\documentclass[a4paper,11pt]{article}
\pdfoutput=1

\usepackage[utf8]{inputenc} 
\usepackage{amsmath}
\usepackage{empheq}
\usepackage{mathrsfs}
\usepackage{amsfonts}
\usepackage{amssymb}
\usepackage{graphicx}
\usepackage{subfigure} 
\usepackage{shadow}
\usepackage{epsfig}
\usepackage{esint}
\usepackage{enumerate}
\usepackage{url}
\usepackage[usenames]{color}
\usepackage{geometry}
\usepackage{fullpage}

\usepackage{authblk}

\begin{document}


\title{Comment on ``Backflow in relativistic wave equations''}

\author[1]{Maximilien Barbier}
\author[2]{Christopher J. Fewster}
\author[3]{Arseni Goussev}
\author[1]{Gregory Morozov}
\author[4,5]{Shashi C. L. Srivastava}

\affil[1]{\small Scottish Universities Physics Alliance, Institute of Thin Films, Sensors and Imaging, University of the West of Scotland, Paisley PA1 2BE, Scotland, UK}

\affil[2]{Department of Mathematics, University of York, York YO10 5DD, UK}

\affil[3]{School of Mathematics and Physics, University of Portsmouth, Portsmouth PO1 3HF, UK}

\affil[4]{Variable Energy Cyclotron Centre, 1/AF Bidhannagar, Kolkata 700064, India}

\affil[5]{Homi Bhabha National Institute, Training School Complex, Anushaktinagar, Mumbai - 400094, India}

\date{}
\maketitle

The stated goal of Ref.~\cite{BBA22Backflow} is to argue that  ``the phenomenon of backflow is a common feature of various physical system[s], quantum and classical, described by the linear wave equations''. The authors challenge a perceived view that ``backflow is inseparably connected with quantum theory''. In particular, the authors strongly oppose statements in the literature that backflow is a ``peculiar quantum effect'', an ``intriguing quantum-mechanical phenomenon'', a ``surprising and clearly nonclassical effect'', a ``classically impossible phenomenon'', and a ``generic purely quantum phenomenon''.

It appears that the authors of~\cite{BBA22Backflow} have 
misunderstood the purpose of the statements to which they object.
To explain, we must distinguish
two mathematically inseparable, but physically distinct phenomena: backflow (B) and quantum backflow (QB). In more detail:
\begin{enumerate}
	
	\item[(B)] {\bf Backflow.} This is a general wave phenomenon, which may be defined, quoting~\cite{BBA22Backflow}, as ``the counterintuitive behavior of the flow of some quantity (energy, probability, etc). Namely, in some regions of space the direction of the flow is {\it opposite} with respect to the direction of all its constituent elementary waves.'' As the authors correctly state, B is ``a common feature of various physical system[s], quantum and classical, described by the linear wave equations''~\cite{BBA22Backflow}. This statement is not subject to doubt and, to our knowledge, has never been challenged.

	\item[(QB)] {\bf Quantum backflow}~\cite{Allcock:1969iii,BrackenMelloy:1994}. This is the phenomenon of backflow specific to quantum particles such as electrons, which concerns the following question: {\it Can the position probability density of a particle flow to the left if the particle's momentum points to the right?}
	Consider a free quantum particle moving in one dimension and suppose that, in the initial state, the
	measured value of momentum would be positive with probability $1$. 
	Is it possible that the probability of finding the particle in the left-hand half-line is higher at a later time than it is initially?
	In classical particle mechanics, the answer to this question is negative. In quantum mechanics, it is positive. This is why numerous authors have referred to QB as a ``peculiar quantum	effect'', ``intriguing quantum-mechanical phenomenon'', ``surprising and clearly nonclassical effect'', ``classically impossible phenomenon'', and ``generic purely quantum phenomenon''. Such phrases concern the 
	contrast between the classical and quantum dynamics of \emph{particles}.
\end{enumerate}

QB arises, of course, because quantum mechanics describes particle mechanics using a wave equation. For exactly the same reason, quantum particles exhibit diffraction and interference phenomena, generic 
behaviour for waves that is forbidden to classical particles. Recall
Feynman's famous contrast of experiments with bullets, waves and electrons, in which he describes the observed behaviour of
electrons as ``impossible, {\it absolutely} impossible, to explain in any classical way, and which has in it the heart of quantum mechanics'' \cite{Feynman}. (Here, Feynman means that it is impossible to account for this behaviour from the perspective of classical particle dynamics. 
Once particle-wave duality is accepted, classical wave dynamics immediately provides the required intuition.)
The statements about QB to which~\cite{BBA22Backflow}
objects are entirely parallel statements that QB also touches the
heart of quantum mechanics. They are not assertions that there are no
other backflow phenomena in physics.

To close, let us emphasise that we do not question the validity of the examples of B reported in Ref.~\cite{BBA22Backflow}. However, we point out that the discussion of backflow for the Dirac equation, presented in section 2 of Ref.~\cite{BBA22Backflow}, is not new. There are several papers on the subject \cite{MB98Probability, SC18Quantum, ALS19Relativistic, PPR}, none of which have been mentioned in Ref.~\cite{BBA22Backflow}.


\bibliographystyle{unsrt}

\end{document}